\documentclass[conference,10pt,top=0.7in,bottom=0.95in]{IEEEtran}
\IEEEoverridecommandlockouts

\usepackage[sort,compress]{cite}
\usepackage{amsmath, amsthm, amsfonts, amssymb, amsbsy,nccmath}

\usepackage{graphicx,color}
\usepackage{textcomp}
\usepackage{mathbbol}
\def\BibTeX{{\rm B\kern-.05em{\sc i\kern-.025em b}\kern-.08em
    T\kern-.1667em\lower.7ex\hbox{E}\kern-.125emX}}

\usepackage{amsmath, amsthm, amsfonts, amssymb, amsbsy,nccmath}
\usepackage{algpseudocode}
\usepackage{balance}
\usepackage{algorithm}
\usepackage{enumerate}
\usepackage{lipsum}
\usepackage{textgreek}
\usepackage{textcomp}
\usepackage[sort,compress]{cite}
\usepackage{epsfig}
\usepackage{epstopdf}
\usepackage{mathtools}
\usepackage{dsfont}
\usepackage{stfloats}
\usepackage{sidecap, caption}
\usepackage{subcaption}
\usepackage{cleveref}

\newtheorem{manualtheoreminner}{Definition}

\theoremstyle{definition}

\theoremstyle{proposition}

\theoremstyle{lemma}
\newtheorem{lemma}{Lemma}

\usepackage[inline]{enumitem}   
\makeatletter
\newcommand{\inlineitem}[1][]{%
\ifnum\enit@type=\tw@
    {\descriptionlabel{#1}}
  \hspace{\labelsep}%
\else
  \ifnum\enit@type=\z@
       \refstepcounter{\@listctr}\fi
    \quad\@itemlabel\hspace{\labelsep}%
\fi}
\makeatother

\newcommand\norm[1]{\left\lVert#1\right\rVert}

\newcommand{\beq}{\begin{equation}}
\newcommand{\eeq}{\end{equation}}

\newcommand{\mD}{\mbox{$\mathcal D$}}

\newcommand{\mC}{\mathcal{C}}

\newcommand{\mO}{\mathcal{O}}
\newcommand{\mF}{\mathcal{F}}
\newcommand{\mN}{\mathcal{N}}
\newcommand{\mA}{\mathcal{A}}
\newcommand{\mE}{\mathcal{E}}

\newcommand{\mW}{\mathcal{W}}

\newcommand{\mT}{\mathcal{T}}

\newcommand{\mU}{{\mathcal U}}

\newcommand{\bmd}
{\boldsymbol{d}}

\def\adots{\mathinner{\mskip0mu\raise0pt\vbox{\kern7pt\hbox{.}}\mskip3mu
          \raise4pt\hbox{.}\mskip3mu\raise8pt\hbox{.}\mskip0mu}}

\DeclareMathOperator*{\argmax}{arg\,max}

\usepackage{bm}

\newcommand{\bmy}{{\bm y}}

\newcommand{\bma}{{\bm a}}

\newcommand{\bGamma}{\boldsymbol{\Gamma}}

\newcommand{\bgamma}{\boldsymbol{\gamma}}

\newcommand{\bpsi}{\boldsymbol{\psi}}

\makeatletter
\newcommand\fs@spaceruled{\def\@fs@cfont{\bfseries}\let\@fs@capt\floatc@ruled
  \def\@fs@pre{\vspace{0.5\baselineskip}\hrule height.8pt depth0pt \kern2pt}%
  \def\@fs@post{\kern1pt\hrule\relax}%
  \def\@fs@mid{\kern2pt\hrule\kern2pt}%
  \let\@fs@iftopcapt\iftrue}
\makeatother

\newcommand{\bit}{\begin{itemize}}
\newcommand{\eit}{\end{itemize}}

\newcommand{\mK}{\mathcal{K}}
\newcommand{\mL}{\mathcal{L}}
\newcommand{\mG}{\mathcal{G}}

\newcommand{\bphi}{\boldsymbol{\phi}}

\DeclarePairedDelimiter\abs{\lvert}{\rvert}%

\usepackage{lipsum}

\restylefloat{algorithm}

\begin{document}

\title{On the Computing and Communication Tradeoff in Reasoning-Based Multi-User Semantic Communications}
\author{\fontsize{1}{1}\selectfont
\IEEEauthorblockN{\fontsize{11}{12}\selectfont Nitisha Singh\IEEEauthorrefmark{1}, Christo Kurisummoottil Thomas\IEEEauthorrefmark{1}, Walid Saad\IEEEauthorrefmark{1}, and Emilio Calvanese Strinati\IEEEauthorrefmark{2}}\IEEEauthorrefmark{1}Bradley Department of Electrical and Computer Engineering, Virginia Tech, Arlington, VA, USA, \\
\IEEEauthorrefmark{2} CEA-Leti, Grenoble, France,\\ \fontsize{8}{8}Emails: \{nitishas,christokt,walids\}@vt.edu, emilio.calvanese-strinati@cea.fr}
\maketitle
\begin{abstract}
Semantic communication (SC) is recognized as a promising approach for enabling reliable communication with minimal data transfer while maintaining seamless connectivity for a group of wireless users. Unlocking the advantages of SC for multi-user cases requires revisiting how communication and computing resources are allocated. This reassessment should consider the reasoning abilities of end-users, enabling receiving nodes to fill in missing information or anticipate future events more effectively. Yet, state-of-the-art SC systems primarily focus on resource allocation through compression based on semantic relevance, while overlooking the underlying data generation mechanisms and the tradeoff between communications and computing. Thus, they cannot help prevent a disruption in connectivity. In contrast, in this paper, a novel framework for computing and communication resource allocation is proposed that seeks to demonstrate how SC systems with reasoning capabilities at the end nodes can improve reliability in an end-to-end multi-user wireless system with intermittent communication links. Towards this end, a novel reasoning-aware SC system is proposed for enabling users to utilize their local computing resources to reason the representations when the communication links are unavailable. To optimize communication and computing resource allocation in this system, a noncooperative game is formulated among multiple users whose objective is to maximize the effective semantic information (computed as a product of reliability and semantic information) while controlling the number of semantically relevant links that are disrupted. To find a Nash equilibrium of the game, an algorithm based on best response is proposed. Simulation results show that the proposed reasoning-aware SC system results in at least a $16.6\%$ enhancement in throughput and a significant improvement in reliability compared to classical communications systems that do not incorporate reasoning.
\end{abstract}


\maketitle
\section{Introduction}

Recent advances in communication technologies, such as ultra-massive multiple-input-multiple-output and exploring higher frequency bands like terahertz may not be sufficient for meeting the rigorous demands posed by emerging applications in future wireless systems such as extended reality (XR) and the metaverse \cite{HashashArxiv2023}. The development of artificial intelligence (AI)-native wireless systems promises to bridge the gap between conventional wireless technologies and the growing demand from emerging use cases \cite{saad2024artificial}. However, current approaches \cite{ZhangCST2019}, which are data-driven, lack generalizability and explainability, hindering them from meeting stringent requirements such as high throughput and continuous reliability.
Herein, a promising approach for future wireless systems is to use the concept of \emph{semantic communication (SC)} \cite{ChaccourArxiv2022} and \cite{StrinatiComNetworks2021}, that incorporate reasoning-enabled AI models at the wireless devices. Integrating generalizable and interpretable models like causal reasoning and neuro-symbolic AI \cite{ChristoTWCArxiv2022} equips the nodes with reasoning capabilities. Reasoning enables nodes to depend less on the channel and communication resources, allowing for improved capabilities for prediction (of future events), generation (of new data), and inference (of missing information). However, with highly intermittent connectivity, effective coordination among agents in a multi-user SC system becomes a challenge. As users depend more on their computing resources for inference, efficient utilization of communication and computing resources is important for users to have reliable connectivity. 
\vspace{-2.2mm}
\subsection{Related Works}

Despite the recent surge in works on SC, most of them fail to address critical challenges in a multi-user SC system such as allocating computing and communication resources, while considering the reasoning capabilities of the end nodes. The majority of prior art in  SC~\cite{XieTSP2021, LiuISIT2021,Farshbafan2023,SunnurArxiv2022} focuses on extracting data semantics and designing encoder and decoder components. Indeed, only few prior works like \cite{YanGC2022,LiuTCCN2023,ZhaoArxiv2024} have investigated the use of semantics for efficient communication and computation resource allocation in multi-user systems. In particular, in \cite{YanGC2022} and \cite{LiuTCCN2023}, the authors restricted their analysis to the optimization of traditional physical layer functions, such as channel assignment, power allocation, and transmit symbols, using semantics-based metrics. 
Even though the authors in \cite{Nguyenmulti} took into account the limited computing and communication resources of the system,  they limit their approach to data-driven AI solutions (in particular, transformer-based) that do not incorporate reasoning at the end-user level. Without reasoning, those AI solutions cannot perform inference on any missing observations stemming from unavailable links. Furthermore, the existing solutions in \cite{YanGC2022, LiuTCCN2023, ZhaoArxiv2024, Nguyenmulti} are centralized, and assume that all the users have perfect knowledge about other user tasks. Additionally, these schemes often necessitate joint training of deep learning modules at both the transmit and receive sides resulting in significant communication overhead.

To overcome the limitations posed by data-driven AI solutions, as highlighted above, and to meet the rigorous performance demands of future wireless systems, integrating reasoning capabilities and thereby optimizing resource allocation is essential. Although \cite{ZhaoArxiv2024} explored the use of symbolic AI techniques to perform multi-user resource allocation, their contribution is limited to traditional MAC layer tasks such as uplink or downlink channel assignment. Moreover, in future connected intelligence systems, the control actions of different users may be correlated due to the shared environment among the users. 
While previous studies like \cite{XieJSAC2022} may have considered interference among multi-user channels, all of the relevant prior works \cite{YanGC2022, LiuTCCN2023, ZhaoArxiv2024, Nguyenmulti, XieJSAC2022}, have overlooked the correlation among the control actions arising from the shared environment. Modeling the communication and control strategies of SC users, while taking into account this correlation is a critical challenge in multi-user SC systems. 
In contrast to \cite{YanGC2022, LiuTCCN2023, ZhaoArxiv2024, Nguyenmulti, XieJSAC2022}, semantic-aware resource allocation must guarantee high semantic reliability in executing user tasks even when communication and computing resources are intermittent.

\subsection{Contributions}
The main contribution of this paper is a novel framework for a multi-user SC system that incorporates reasoning capabilities at end-users to enhance network reliability in scenarios where link availability is intermittent. In contrast to classical systems, where a communication link failure results in a complete interruption, our proposed SC system enables users to leverage AI reasoning techniques and generate the data that would have been received over an unavailable link. In our framework, SC users achieve this by discovering causal relationships in the data, understanding the underlying data-generation mechanism, and leveraging their computing resources for reasoning. 
Furthermore, the correlation among control actions stemming from the shared environment motivates the formulation of a game that captures the dependencies among communication and computing resource allocation decisions. Our proposed scheme is a noncooperative game that learns the optimal resource allocation decisions to maximize communication and control utilities. We propose a sequential best response algorithm that can find a local Nash equilibrium of the game. 
Simulation results illustrate that the proposed SC system achieves an enhancement of at least $16.6\%$ in throughput and a 9-fold increase in reliability compared to classical communication systems lacking reasoning capabilities.

The rest of the paper is organized as follows. Section II presents the system model. Section III describes the problem formulation and solution. Section IV discusses the results and Section V concludes the paper.
\section{System Model}
\begin{figure}[t]
\centerline{\includegraphics[width = \linewidth]{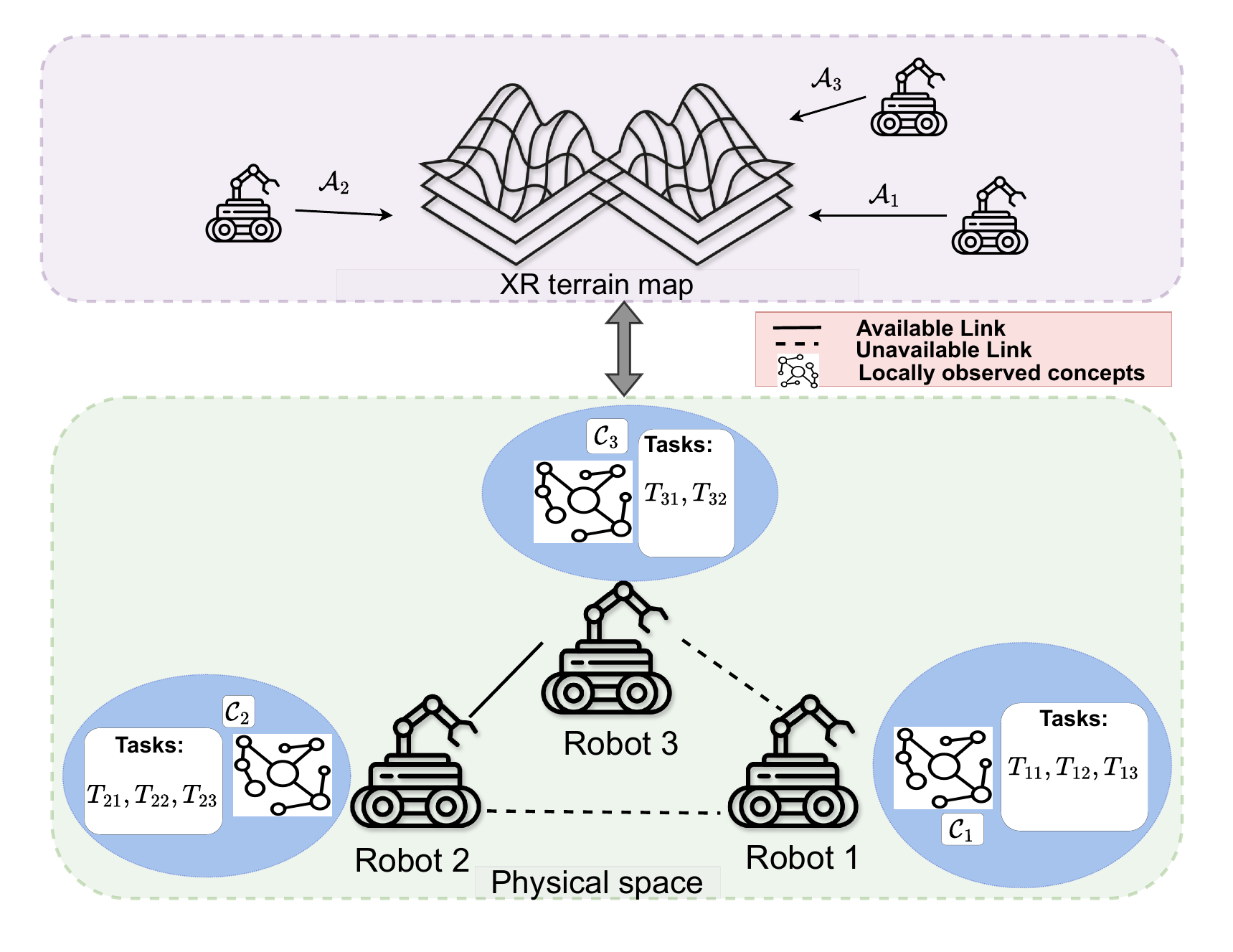}}
    \caption{\small{An illustration of a decentralized robotic terrain mapping system. Robots, operating in a shared physical environment, collaborate to build a virtual environment such as a 3D map. They exchange information through unreliable communication links, relying on local computing resources to reason about missing data. The control actions, represented by $\mA_k$, correspond to the robots' movements and actions. The tasks in the physical space involve identifying and classifying terrain features, hazards, and potential resources. This information is then used to build a shared virtual map, accessible through an XR interface, facilitating human supervision.}}
    \vspace{-5mm}
    \label{XR} 
\end{figure}

 We consider an end-to-end multi-user wireless system composed of a set $\mathcal{K}$ of $K$ users communicating with each other over intermittent links. The unreliability of the links poses a challenge, as any disruption in communication can lead to degraded user experiences, including latency and jitter. Classical communication systems address this unreliability through techniques like efficient beam recovery or ACK/NACK procedures, yet these methods cannot entirely prevent link disruption, and they often do it at the expense of reduced throughput. We consider an SC system
\cite{ChaccourArxiv2022}, in which users can rely on their computing capabilities and reason what could have been transmitted on the unavailable link, thus, maintaining link continuity. We represent the network with a graph $\mL = (\mK, \mE)$, where $\mE = \{(i, j)|i, j \in \mK, i\neq j\}$ is the set of active links between the users in $\mathcal{K}$. We assume a time division duplexing (TDD) system in which a user can act as either a transmitter or a receiver at a given time slot. 
A prime example of such a system is a group of collaborative robots navigating and mapping a complex terrain, as illustrated in Fig \ref{XR}. Each robot uses multi-modal sensors (e.g., cameras, LIDAR) to perceive its surroundings, gathering high-dimensional data that may be correlated. They coordinate their actions and share their observations to complete individual tasks (e.g., exploring specific areas, identifying hazards, and building a 3D model), ultimately contributing to a shared mapping objective. 
In this example, the users observe a complex and high-dimensional environment using multi-modal data sensors which may be correlated. We consider a scenario under limited communication and computation resources, as the user data is high-dimensional and possibly correlated, it is more efficient for users to communicate by learning a compact, semantic representation that captures the causal relationships between identified concepts in the data rather than relying on traditional communication methods \cite{ChristoTWCArxiv2022}. Users can employ the causal discovery methods described in \cite{ChristoTWCArxiv2022} to identify features within their observed data that convey relevant meaning for the multi-user tasks. These identified features are referred to as \emph{semantic concepts} and the relationships among them can be captured using a structural causal model (SCM). 
The users are equipped with sensors that capture the environment the user moves in. Due to the limited field of view of the sensors at user $k$, each user can only observe a $d$-dimensional local view of the global set of semantic concepts $\mC$ using its SCM. This representation is captured using the observation mapping $O_k:\mC \rightarrow \mO_k \subseteq \mathbb{R}^d$, 
where the local observation $O_k$ at user $k$ can be described by the causal relationships among the extracted concepts, $\mC_k = \{c_{k1}, c_{k2}, ..., c_{kd}\}$. The causal relationships are represented as a graph, where the nodes represent the concepts and the edges represent the relationships between the concepts. The global observation set is defined as $\mO =\bigcup_{k = 1}^K \mO_k$. In our example, the robots have a partial view of the terrain. They identify semantic features of interest from their observations and exchange them with each other to collaboratively complete a team objective (building a virtual 3D map), which can be divided into individual tasks. User $k$ has a task $T_k \in \mT_k$ that requires other users’ observations to execute. Each task $T_k$ is represented by a tuple $(\mW_k, \mF_k, \mA_k)$, where $\mW_k$ is the set of semantic concepts required for the completion of the task and $\mF_k$ contains the number of computation cycles per second required to reason the concepts in $\mW_k$. $\mA_k$ is the set of control actions $a_k$ that user $k$ can take, and its ideal control policy is defined by the distribution $p(a_k \mid \mC, \bma_{-k})$ respectively, where $\bma_{-k}$ is the action vector of users other than $k$, i.e., the users in set  $\mathcal{K} \setminus \{k\}$. The ideal control policy requires user $k$ to know the global concepts $\mC$, which necessitates communication between users. The control actions that a user takes depend on the quality of inference of the concepts and the control actions of other users as the actions affect the user's shared environment. 
\subsection{Computing and Communication Model}
The connected neighborhood of a user $k$ is defined by $\mN_k = \{j \in \mK| (k, j) \in \mE, P_{kj} > 0\}$ when a link exists between users $k$ and $j$ and when the link to user $j$ is semantically important to user $k$. Here, $P_{kj}$, represents the semantic importance as the fraction of concepts that user $k$ needs from user $j$ in order to complete its task. The semantic representations received by user $k$ from its neighbors are $\widehat{\mC}_k = \{\widehat{m}^{(k)}_{jr}\mid j \in \mN_k, r\in \left[1, d\right]\}$ and the representations that user $k$ transmits to its neighbors are $\mC_k = \{m_{kr}^{(j)}| j \in \mN_k\}, r\in \left[1,d\right]$, where $m_{kr}^{(j)}$ is the semantic representation corresponding to $c_{kr}$. The representation received by user $k$,  
 $\widehat{m}_{jr}^{(k)} \in \widehat{\mC}_k$ is distorted due to fading and interference effects of the communication link. The  distribution $p(\widehat{m}^{(k)}_{jr}\mid m^{(k)}_{jr})$, represents the likelihood of receiving $\widehat{m}_{jr}^{(k)}$ given that $m_{jr}^{(k)}$ was sent, thus capturing the communication link quality. 
If the received information cannot be decoded, then the users must depend on computing resources to deduce missing information. For a given user $k\in \mathcal{K}$, this is captured by the decision vector $\bmd_k$, with each scalar $d_{ik}$ assuming a discrete value corresponding to the decision for link $i$. If a user is transmitting on a link with user $i$ then $d_{ik} = 0$. If user $k$ is receiving on that link and the reconstructed signal is of acceptable quality then $d_{ik} = 1$. Meanwhile, if user $k$ is using its computing resources to understand what user $i$ intended to send then $d_{ik} = 2$. User $k$ can also decide to do nothing on link $(i,k)$ in which case $d_{ik} = 3$. At any given communication instance, a user acts either as a transmitting or receiving link. This distinction is intuitive since it is conceivable that only a subset of users may possess observations relevant to the execution of tasks that must be transmitted. If a user requires a semantic concept to accomplish its goal, which is absent from its own observations $\mO_k$ or in its reconstructed concepts $\widehat{\mC}_k$, then this concept must come from an unavailable link. In such a scenario, the user may allocate its computing resources to reason the concept. Since edge users are often resource-constrained, each user will have a maximum computing capacity of $f^{\textrm{max}}_k$ cycles per second. This constrains the computing resources that can be allocated to the unavailable links. If computing resources are not available, the user must wait until the computing resources are freed or the wireless link becomes available. Since downtime is undesirable, all users cooperate to minimize it while completing their goals. 
The semantic reliability of communication on the link from user $k$ to user $j$ is given by the following probability: $r_{kj} = P\left(\frac{1}{\abs{\mW_j}}\sum\limits_{r} \mathbb{1}(c_{kr}^{(j)} \in \mW_j) E_c[m_{kr}^{(j)}, \hat{m}_{kr}^{(j)}] < \delta\right)$, where $E_c[m_{kr}^{(j)}, \widetilde{m}_{kr}^{(j)}] = \norm{m_{kr}^{(j)}- \widetilde{m}_{kr}^{(j)}}^2$ represents the semantic distortion and $\mathbb{1}$ is the indicator function. The value of $\delta$ depends on the perceptual quality of the actions performed by the user ($\alpha_k$) and can be defined as the minimum value of the error $E_c$ that can be tolerated such that $\mathrm{KL}\left(p(a_k\mid \mC) || p(a_k\mid\widehat{\mC}_k)\right) \leq \alpha_k$. On the transmission link, each user seeks to maximize the semantic impact generated by a transmission on the link ($k, j$), defined as $\eta_{kj}$. Simultaneously, the transmitter aims to ensure that the semantic reliability on the link will be as close to one as possible, i.e., we must have $r_{kj} \geq 1 - \epsilon$, where $\epsilon$ is arbitrarily small. Here, $\eta_{kj}= \frac{1}{d}\sum\limits_{r=1}^d \frac{I(c_{kr};m_{kr}^{(j)})}{l_{kr}^{(j)}}$ is the ratio of the number of bits required to transmit the representation $m_{kr}^{(j)}$ in the classical sense (without semantic communication) to the number of bits needed to transmit the semantic representation. The mutual information conveyed by $m_{kr}^{(j)}$ about $c_{kr}$, ${I}(c_{kr};m_{kr}^{(j)})$ is the number of bits to be allocated to the representation $m_{kr}^{(j)}$ when transmitted in the classical sense. The number of bits allocated to the semantic representation $m_{kr}^{(j)}$ is $l_{kr}^{(j)}$. 
Note that $\eta_{kj} \in \left[1,\infty\right]$. Clearly, the semantic impact is higher when we allocate fewer bits, and lower ($=1$) when $l_{kr}^{(j)} = I(c_{kr};m_{kr}^{(j)})$. Next, we look at how to perform reasoning when the communication links are unavailable.
\subsection{Reasoning Model and Semantic Information}
When $d_{jk} = 2$, user $k$ reasons any concept $m_{jr}^{(k)} \in \widehat{\mC}_{k}$ by using the concept of interventions from causal reasoning~\cite{BareinboimACM2020}. Here, an intervention is analytically formulated as:
\beq
\begin{aligned}
\widetilde{m}_{jr}^{(k)} = \argmax\limits_{m_{jr}^{(k),0} \in \widehat{\mC}_k} p(\widehat{\mC}_k  \backslash m_{jr}^{(k)}\mid do(m_{jr}^{(k)}) = m_{jr}^{(k),0}),
\end{aligned}
\label{eq_interventions}
\eeq
\noindent{which can be explained as user $k$ inferring the semantic concept $m_{jr}^{(k)}$ that best explains the remaining concepts received, defined as $\widehat{\mC}_k  \backslash m_{jr}^{(k)}$. The operation $do(m=m^{(0)})$ entails removing all incoming edges to $m$ (in the causal graph formed by the semantic concepts) and then setting it to a specific value $m^{(0)}$. The resulting reasoning distortion can be written as $E_r[m_{jr}^{(k)}, \widetilde{m}_{jr}^{(k)}] = \norm{m_{jr}^{(k)}- \widetilde{m}_{jr}^{(k)}}^2$. As such, we can capture the reasoning reliability probabilistically as $b_{jk}=P(\frac{1}{\abs{\mW_j}}\sum\limits_r \mathbb{1}(c_{kr}^{(j)}\in \mW_k)(E_r[m_{jr}^{(k)}, \widetilde{m}_{jr}^{(k)}] \leq \delta)$. Next, we obtain the semantic information as follows.}
\begin{lemma}
The semantic information extracted at any user $k$ can be written as the sum of the information extracted from the available communication link and that obtained via reasoning, and can be written as follows:
\beq
\mathbb{S}_{k} = \sum\limits_{j}\mathbb{1}(d_{jk}==1)\mathbb{S}_{(j,k),c} + \sum\limits_{j}\mathbb{1}(d_{jk}==2)\mathbb{S}_{(j,k),r},
\label{eq_S_k}
\eeq
where $\mathbb{S}_{(j,k),c}$ is written as \eqref{eq_Sjk_c} and $\mathbb{S}_{(j,k),r}$ as \eqref{eq_Sjk_r}.
\end{lemma}
\begin{IEEEproof}
For any link ($j, k)$, 
$\mathbb{S}_{(j,k),c}$ can be written as the traditional mutual information between the transmitted representation $m_{jr}^{(k)}$, and the received representation $\widehat{m}_{jr}^{(k)}$, i.e.,  
\begin{equation}
    \mathbb{S}_{(j,k),c} = \sum\limits_{r} p(m_{jr}^{(k)}\mid \widehat{m}_{jr}^{(k)}) \log_2\left(\frac{p(m_{jr}^{(k)}\mid\widehat{m}_{jr}^{(k)})}{p(m_{jr}^{(k)})}\right),
    \label{eq_Sjk_c}
\vspace{-0mm}\end{equation}
and $\mathbb{S}_{(j,k),r}$ can be written using the interventional distribution \eqref{eq_interventions}:
\begin{equation}
\begin{aligned}
       \mathbb{S}_{(j,k),r} = &\sum\limits_rp(\widehat{\mC}_k  \backslash m_{jr}^{(k)}\mid \textrm{do}(m_{jr}^{(k)}) = m_{jr}^{(k),0})\\ &\log_2\left(\frac{p(\widehat{\mC}_k  \backslash m_{jr}^{(k)}\mid \textrm{do}(m_{jr}^{(k)}) = m_{jr}^{(k),0})}{p(\widehat{\mC}_k  \backslash m_{jr}^{(k)})}\right).
    \label{eq_Sjk_r}
\end{aligned}
\end{equation}
Since the extracted semantic information results from either communication or reasoning, we can write \eqref{eq_S_k}.
\end{IEEEproof}
When $d_{jk} = 1$, each user intends to maximize the information received that is useful to its task, which can be interpreted as the product of reliability and semantic information communicated, i.e., $r_{jk}\mathbb{S}_{(j,k),c}$. Similarly, when $d_{jk} = 2$, the resulting useful information can be written as $r_{jk}\mathbb{S}_{(j,k),r}$. 
\vspace{-4mm}
\subsection{Communication and Control Utility}
\vspace{-2mm}
After a user has chosen their decision $\bmd_k \in \mD_k$ and performed a control action $a_k$, they receive a utility based on the function $U_k: \mD_k \rightarrow \mathbb{R}$, which is a combination of control and communication utilities. 
We define a communication utility $U_k^c(d_k)$ that captures the effective semantic information reconstructed when the user acts as a receiver, as well as the effective semantic impact when it is a transmitter, as follows:
\beq
\begin{aligned}
 &U_k^c(d_k) = 
      \underbrace{\sum_{j} \mathbb{1}(d_{kj} = 0)r_{kj}\eta_{kj}}_{\text{effective semantic impact}} \\&+ \underbrace{\sum_j \mathbb{1}(d_{kj} = 1)r_{jk} \mathbb{S}_{(j,k),c}  +  
       \sum_j \mathbb{1}(d_{kj} = 2)b_{jk} \mathbb{S}_{(j,k), r}}_{\text{effective}\,\, \text{semantic info}}.
       \end{aligned}
\eeq
\begin{figure*}
\beq
\begin{aligned}
    U_{k,j}^r(a_k, \bm a_{-k}) &= \sum\limits_{j\in \mN_k} \mathbb{1}(d_{kj} = 1)\left(\left(\frac{2}{1+e^{-\zeta(a_k,a_j)}}-1\right) - \frac{\theta_1}{K}\right)  + \sum\limits_{j\in \mN_k}\mathbb{1}(d_{kj} = 2)\left(\left(\frac{2}{1+e^{-\zeta(a_k,a_j)}}-1\right) - \frac{\theta_2}{K}\right) \\ &+
    \sum\limits_{j\in \mN_k}\mathbb{1}(d_{kj} = 3)\left(\left(\frac{2}{1+e^{-\zeta(a_k,a_j)}}-1\right) - \frac{\theta_3}{K}\right)
    \label{eq_gene_actions}
\end{aligned}
\eeq
\vspace{-8mm}
\end{figure*}
We consider any user $k$'s control actions to be a function of the communication and reasoning accuracies. When both fail, user $k$ takes random actions. We write the actions as $a_k=\zeta(d_k,\mC_k,\widehat{\mC}_{k})$, where
\begin{equation}\label{controlact}
\begin{aligned}
\zeta(d_k,\mC_k,\widehat{\mC}_{k}) = \sum_{j} \mathbb{1}(d_{kj}=1) \sum_{r} f(E_c[m_{kr}^{(j)},\widehat{m}_{kr}^{(j)}])\\+ \sum_{j} \mathbb{1}(d_{kj}=2)\sum_{r} f(E_r[m_{kr}^{(j)},\widetilde{m}_{kr}^{(j)}]) + \sum_{j} \mathbb{1}(d_{kj}=3) v_{k},
\end{aligned}
\end{equation}
where $v_{k}$ is uniformly random sampled from $\mA_k$. Here, $f$ can be a non-linear function that needs to be learned. If the users coordinating with $k$ are aligned in terms of their actions, then the task difficulty reduces by an amount of $\frac{\theta_i}{K}$, where $\theta_i$ is a constant. The alignment between actions $a_k$ and $a_j$ can be defined as the cosine similarity, i.e.,  $\zeta(a_k,a_j) = \frac{a_k \cdot  a_j}{||a_k|||| a_j||}$. We define the utility obtained by user $k$ as a sum of the reduction in task difficulties across all coordinating users.
Because the users share a common environment, the utility of each user $k$ depends on the control actions of other users, defined as the vector $\bma_{-k}$. Thus, we define a so-called control utility,
$U_k^r(a_k, \bm a_{-k})$ in \eqref{eq_gene_actions} as a generalized version of the coordination payoff structure in \cite{WeiCISS2023} which is limited to binary actions. 
For notational simplicity, we define the combined communication and control strategy as $\gamma_k = \left[a_k,d_k\right]$. 
Further, combining the two objectives, we write user $k$'s utility function as:
\beq
 \begin{aligned}
  U_k(\gamma_k, \bgamma_{-k}) = \alpha U_k^c(d_k)
 + \beta U_k^{r}(a_k(d_k),\bma_{-k}(\bmd_{-k})).
     \label{eq_utility}
 \end{aligned}
 \eeq
Where $\alpha$ and $\beta$ are weight factors.
Next, we study the communication-computation tradeoff problem in a multi-user SC system.
\section{Problem Formulation and Proposed Solution}
A user's strategy involves learning to communicate representations that convey the desired meaning or concepts accurately with minimal bits. Each user's objective here is to maximize the semantic information reconstructed with as high semantic reliability as possible by optimizing their computing and communication resources. Given that users function within a shared environment, their control actions are interdependent. This means each user, $k$, is affected by the actions taken by other users, which depends upon the reconstruction quality of semantic concepts across all users. This indicates that the utility of user $k$ depends not only on $d_k$ but also on the decisions of other users, represented as $\bmd_{-k}$. A promising way to formulate the resulting optimization problem is as a multi-user noncooperative game \cite{Han_Niyato_Saad_Başar_2019}. Here, a game-theoretic approach is apropos because of the intricate interdependencies among the user computing and communication decisions.
The game can be described in strategic form as the tuple $\mG = \left(\mK,\mD,\mA,\mU\right)$, where $\mD$ is the set of all decision vectors $\bmd_k$, $\mA$ is the set of control actions, and $\mU$ is the set of all utilities $U_k$. 
Given $\bmd_{-k}$ and $\bma_{-k}$, the goal of each user $k \in \mathcal{K}$ is to solve the following constrained optimization problem: 
\begin{subequations}\label{prob}
\begin{align}
& \left[\gamma_k^{\ast}\right] \in \arg\max_{\gamma_k} U_k (\gamma_k, \bgamma_{- \bm k}^{\ast}) \label{eq:objective} \\
\textrm{s.t.}, & \quad \sum_{j \neq k} \mathbb{1}(d_{kj} = 2)f_{kj} \leq f_k^{\textrm{max}} \label{eq:constraint1},  \forall k\\
& \sum_k\sum_j \mathbb{1}(d_{kj} = 0)\sum_{r}\mathbb{1}(c_{jr}\in \mW_k)l_{jr}^{(k)} \leq N, \label{eq:constraint2} \\
& \sum_{j \neq k} \mathbb{1}(d_{kj} = 3) P_{k,j} \leq L_{\textrm{max}},
\label{eq:constraint3}
\end{align}
\end{subequations}
where \eqref{eq:constraint1} represents the computing constraints at every user. In other words, the computing resources allocated to each link $f_{kj}$ should not exceed $f_k^{\textrm{max}}$. \eqref{eq:constraint2} represents the communication constraints, where the total number of bits transmitted across all users must remain less than $N$, and \eqref{eq:constraint3} constrains the number of semantically relevant links that can be inactive at a user should be less than $L_{\textrm{max}}$.  The studied game is a static noncooperative game and, hence, one suitable solution is the concept of a Nash equilibrium defined next:
\begin{manualtheoreminner} 
    A strategy profile $\bGamma = \left(\gamma_1,\cdots,\gamma_K\right)$ is a \emph{Nash equilibrium} of the multi-user game $\mG$ if
    \beq
    \begin{aligned}
\vspace{-1mm}U_k(\gamma_k^{\ast},\bgamma_{-k}^{\ast}) \geq U_k(\gamma_k,\bgamma_{-k}^{\ast}),  \gamma_k \in \bGamma, \forall i \in \left[K\right],
    \label{eq_NE}
    \end{aligned}
    \eeq
    where $\gamma_k = \{d_k,a_{k}\}$.
\end{manualtheoreminner}
In our system, a Nash equilibrium corresponds to the optimal control and communication actions ($\gamma_k, \bm\gamma_{-k})$ that maximizes user $k$'s utility $U_k(\gamma_k,\bm\gamma_{-k})$. The equilibrium allows us to compute the optimal communication and computing resource allocation decision in a decentralized manner without the need to know how other users execute these decisions or how the reconstruction quality of semantic concepts influences their control actions.
\subsection{Proposed Solution}
In our system, the objective of user $k$ can be considered as performing the optimal intervention ($a_k$) such that its control utility is maximized. This can be captured as a causal graph, with the directed relations $\{\mC, \bma_{-k} \} \rightarrow a_k \rightarrow U_k^r $. Here, the interventions can be modeled as the mean of a Gaussian process (GP), inspired by causal Bayesian optimization \cite{AgliettiPMLR2020}. $f$ represents the mean of a GP, mapping the semantic distortion to the values of the actions. We particularly choose GP since it can learn any non-linear function as the mean of the parameterized Gaussian distribution.
For analytical simplicity, we assume that user $k$ observes the actions of others, $\bma_{-k}$. Here, the observations at any user $k$, are defined as the vector $\bmy_k =\left[\widehat{\mC}_k,\bma_{-k}\right]$, which includes the reconstructed semantics and the actions of coordinating users. Given $\bmy_k$, user $k$ objective is to compute the Nash equilibrium solution \eqref{eq_NE}. We propose to optimize $U_k(\gamma_k,\bgamma_{-k})$ using stochastic gradient descent (SGD). As such, we parameterize the decision vectors $d_k$ with a non-linear structure, which is a promising way to represent any non-linear decision vector $d_k$ as defined below.
\begin{manualtheoreminner}
Given the observations $\bmy_k$, the \emph{decision vector $d_k$} of user $k$ has the following non-linear structure:
\beq
\begin{aligned}\label{affinestrat}
\bmd_{k}(\bphi_k) =   \vspace{-2mm}\begin{cases}
      1 & \text{if}\,\,\, g_{k,\bpsi}(\bphi_k^T\bmy_k) \leq \tau_k^1  \\
      2 & \text{if}\,\,\, g_{k,\bpsi}(\bphi_k^T\bmy_k) \in \left[\tau_k^1,\tau_k^2\right]\\
      3 & \text{otherwise}
    \end{cases} 
\end{aligned}
\eeq
where $\bm\phi_k$ parameterizes the policy and $\tau_k^1$, $\tau_k^2$ are the threshold values for taking a decision. $g_{k,\bpsi}$ is the parameterized non-linear vector function, where $\bpsi$ are to be learned.
\end{manualtheoreminner}
The policy described in \eqref{affinestrat} can be implemented using a cascade of linear neural network layers with a non-linear activation function. The parameters $\bm\phi_k$ can be learned by using SGD. To learn $\bm\phi_k$ that maximizes the utility for user $k$, we set the loss function equal to the negative of the total utility of user $k$.
\setlength{\textfloatsep}{0pt}\begin{algorithm}[t]
\caption{\small Proposed algorithm based on sequential best response}
\label{algo:consecutive_best_response}
\begin{algorithmic}[1]\small
\State \textbf{Given:} $K$, $p$, randomly initialize $\mE$, $\mathcal{C}_k$, $\mathcal{T}_k$\;
\For{$i \gets 1$ to $N_t$}
     \State Agent $k$ chooses a decision according to \eqref{affinestrat}\;
   \State Control actions for user $k$ are computed according to \eqref{controlact}\;
     \State Utility for user $k$ is computed according to \eqref{eq_utility}\;
    \State Update loss for $\bm\phi_k$ using SGD\;
    \State Randomly update $\mE$ (link availability)\;
\EndFor
\end{algorithmic}
\end{algorithm}
The best response of user $k$ is determined by the inequalities \cref{eq_bestresponse1,eq_bestresponse2,eq_bestresponse3}. For the link from $j$ to $k$, the user decides to consider the communicated information for its task, if the inequalities \eqref{eq_bestresponse1} and \eqref{eq_bestresponse2} hold.
\begin{figure*}
\beq
\begin{aligned}
     &r_{jk} \mathbb{S}_{(j,k),c} + \left(\frac{2}{1+e^{-\zeta(a_k,a_j)}}-1\right) - \frac{\theta_1}{K}   \stackrel{d_{kj}=1}{\geq}  b_{jk} \mathbb{S}_{(j,k),r} + \left(\frac{2}{1+e^{-\zeta(a_k,a_j)}}-1\right) - \frac{\theta_2}{K}
     \end{aligned}\label{eq_bestresponse1}
     \eeq
     \beq
\begin{aligned}
&r_{jk} \mathbb{S}_{(j,k),c}+\left(\frac{2}{1+e^{-\zeta(a_k,a_j)}}-1\right) - \frac{\theta_1}{K}  \stackrel{d_{kj}=1}{\geq}
    \left(\frac{2}{1+e^{-\zeta(a_k,a_j)}}-1\right) - \frac{\theta_3}{K}.
\end{aligned}\label{eq_bestresponse2}
\eeq
\end{figure*}
Otherwise, the user decides to rely on reasoning, if \eqref{eq_bestresponse3} holds.
\begin{figure*}
\beq
\begin{aligned}
    &b_{jk} \mathbb{S}_{(j,k),r} + \left(\frac{2}{1+e^{-\zeta(a_k,a_j)}}-1\right) - \frac{\theta_2}{K}  \stackrel{d_{kj}=2}{\geq}
    \left(\frac{2}{1+e^{-\zeta(a_k,a_j)}}-1\right) - \frac{\theta_3}{K}
\end{aligned}
\label{eq_bestresponse3}
\eeq
\vspace{-7mm}
\end{figure*}

\begin{figure}[htbp!]
    \centering
\includegraphics[width = 8cm]{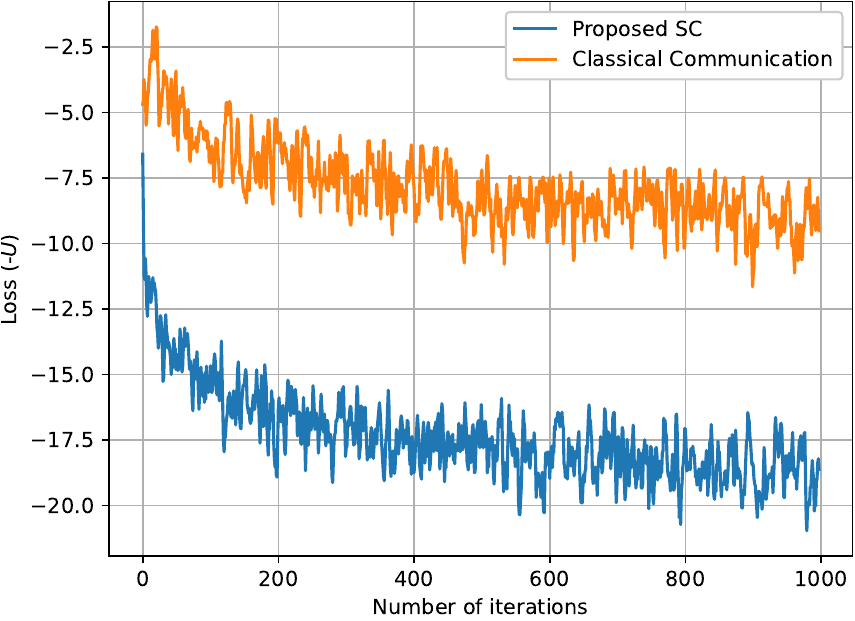}
    \caption{ Loss vs iterations for $p=0.5$}
    \label{Convergence}
\end{figure}

The objective \eqref{eq_utility} is a non-convex function of the strategies $\bmd_k$. We propose Algorithm \ref{algo:consecutive_best_response} based on sequential best response to obtain the optimal strategy $\bmd_k$. Since we rely on best response, once the algorithm converges, we will reach a Nash equilibrium. However, because of the complexity of the utility functions, it is challenging to prove the algorithm's convergence analytically. We show convergence via numerical results. Fig. \ref{Convergence} shows that the utility function converges after a few hundred iterations of running Algorithm  \ref{algo:consecutive_best_response}.
\section{Simulation Results and Analysis}
In this section, we evaluate the performance of the proposed reasoning-based SC system. We consider a classical wireless system that does not use semantics as a baseline for our comparisons. For our simulations, we consider a set of 5 users and iteratively compute the best response for each user using Algorithm \ref{algo:consecutive_best_response}. The decision policy $\bmd_{k}(\bm\phi_k)$ is implemented using a single linear layer with a sigmoid activation and its parameters $\bm\phi_k$ are updated by SGD using a loss set to $-U_k$. Assuming the causal discovery of semantic concepts to be known, the observations for every user are sampled from a standard normal distribution.  The availability of the links is modeled using a Bernoulli distribution where the probability $p$ is varied from $0.1$ to $0.9$. The communication link between the users is assumed to be an additive white Gaussian noise (AWGN) channel with a fixed signal to noise ratio (SNR) of 10 dB.
\par
\begin{figure}[htbp!]
    \centering
    \includegraphics[width = 8cm]{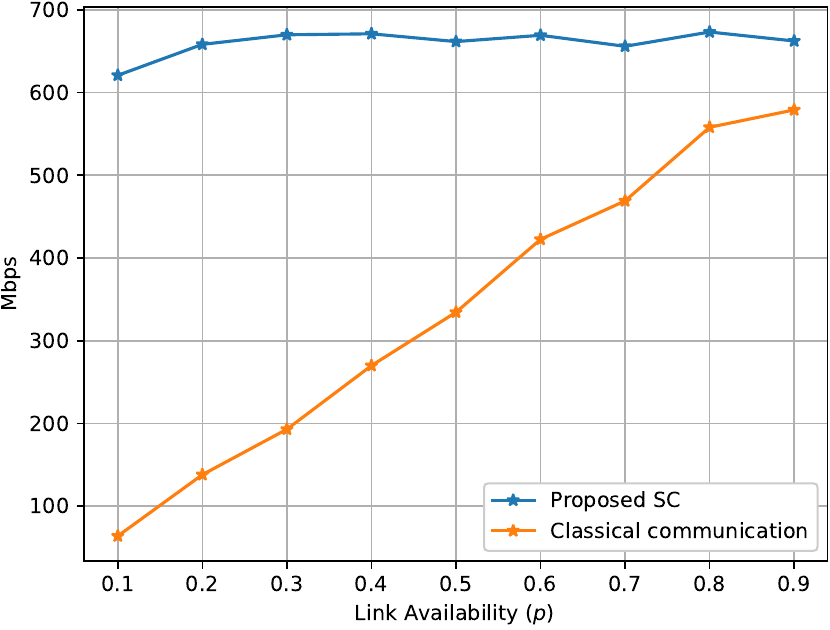}
    \caption{ System throughput as a function of link availability $p$}
    \label{throughput}
\end{figure}

 Fig. \ref{throughput} shows the average throughput of all users resulting from the proposed SC system compared with a classical wireless system, as a function of the link availability probability. From Fig. \ref{throughput}, we observe that the proposed SC system can guarantee a constant throughput even when communication links are scarce, as it can emulate the link via reasoning. In particular, the proposed SC framework achieved a $6$-fold increase in throughput compared to the classical communication system at low link availabilities. At high $p$, the achieved throughput of the proposed SC system is still $16.6 \%$ greater than the throughput achieved by the classical system as transmitting semantics captured via causal reasoning increases the system's efficiency. 
\par
\begin{figure}[htbp!]
    \centering
    \includegraphics[width = 8cm]{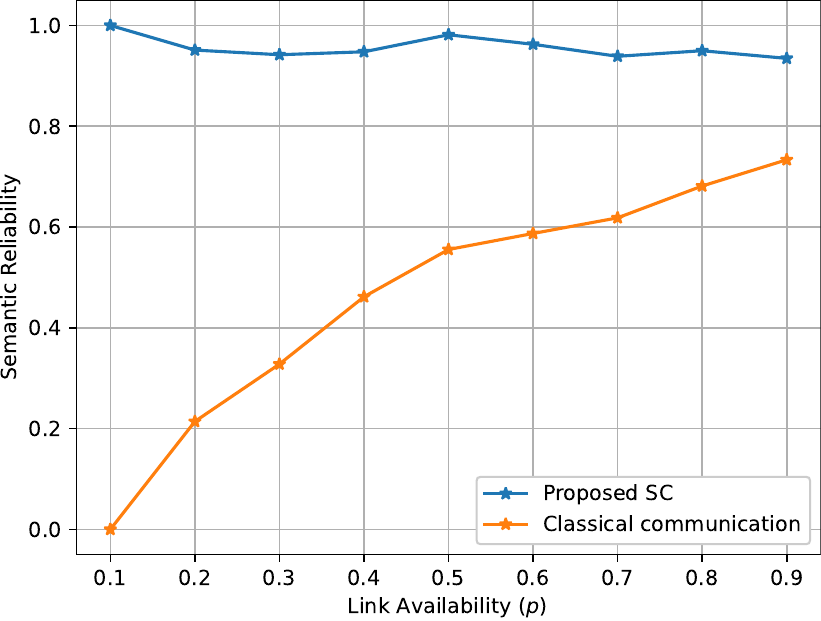}
    \caption{ Semantic reliability as a function of link availability $p$}
    \label{reliability}
\end{figure}

 Fig. \ref{reliability} compares the average semantic reliability of the proposed SC system with that of the classical system, for various values of link availability. We observe that the gap between the proposed system and the baselines becomes smaller as $p$ grows larger. However, for low $p$, the proposed system achieves a reliability that is 9 times higher than that of the baseline. This implies that the proposed SC system can guarantee reliable connectivity via reasoning even when a link does not exist. This serves as a proof of concept of the idea of semantic showers for robust channel control as proposed in \cite{ChaccourArxiv2022}.

\begin{figure}[htbp!]
    \centering \includegraphics[width = 8cm]{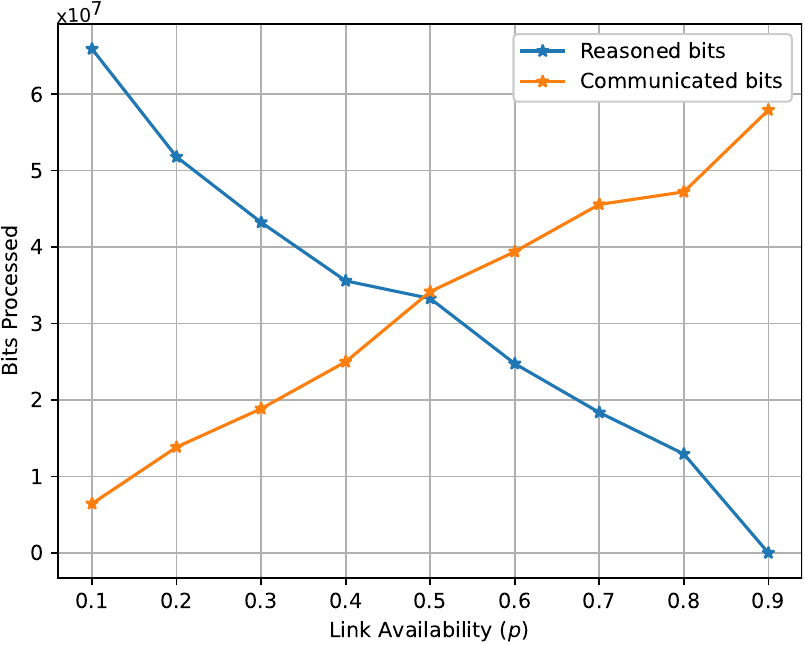}
    \caption{ Reasoned and communicated bits as a function of link availability $p$}
    \label{reasoning}
\end{figure}

Fig. \ref{reasoning} shows the average number of bits a user receives or reasons as a function of the link availability $p$ for the proposed SC system. We observe the tradeoff between computing and communication as $p$ varies. It is evident that when link availability is low, the users have to rely more on reasoning to regenerate content. Conversely, when the link availability is high, the reliance on reasoning decreases as the users can communicate over available links.

\section{Conclusion}
In this paper, we have studied the communication-computing resource allocation problem in a multi-user SC system where users coordinate in a shared environment. In particular, our novel system relies on causal reasoning to maintain reliable connectivity in scenarios with intermittent links. We have formulated a noncooperative game and propose an iterative algorithm based on sequential best response to find a local Nash equilibrium. The equilibrium corresponds to the optimal control and communication actions that maximize the utility in terms of semantic reliability and the effectiveness of coordination between users. Simulation results demonstrate that our proposed SC system achieves better throughput and reliability, even with high link intermittence, when compared to a classical wireless system that does not integrate reasoning into end users. 
\bibliographystyle{IEEEbib}
\bibliography{semantic_refs}
\end{document}